\documentclass[a4paper,11pt]{article}
\pdfoutput=1 

\usepackage{jinstpub} 

\usepackage{lineno}
\modulolinenumbers[5]

\usepackage{graphicx}
\usepackage{amsmath} 
\usepackage{subfigure}
\usepackage{booktabs}
\usepackage{xcolor}
\usepackage{adjustbox}
\usepackage{floatrow}
\newfloatcommand{capbtabbox}{table}[][\FBwidth]
\usepackage{blindtext}

\usepackage{lineno}
\begin{document}

\title{\boldmath Direct detection of charged particles with SiPMs}


\author[a,1]{F. Carnesecchi \note{Corresponding authors.}}
\author[b,1,2]{, G. Vignola \note{Now at Deutsches Elektronen-Synchrotron DESY, Hamburg, Germany}}
\author[b]{, N. Agrawal}
\author[b]{, A. Alici}
\author[c]{, P. Antonioli}
\author[b]{, S. Arcelli}
\author[b]{, F. Bellini}
\author[c]{, D. Cavazza}
\author[b]{, L. Cifarelli}
\author[b]{, M. Colocci}
\author[d]{, S. Durando}
\author[b]{, F. Ercolessi}
\author[e]{, M. Garbini}
\author[b]{, M. Giacalone}
\author[c]{, D. Hatzifotiadou}
\author[a]{, N. Jacazio}
\author[c]{, A. Margotti}
\author[b]{, G. Malfattore}
\author[c]{, R. Nania}
\author[c]{, F. Noferini}
\author[c]{, O. Pinazza}
\author[c]{, R. Preghenella}
\author[f]{, R. Ricci}
\author[c]{, L. Rignanese}
\author[b]{, N. Rubini}
\author[c]{, E. Scapparone}
\author[b]{, G. Scioli}
\author[b]{, S. Strazzi}
\author[c]{, S. Tripathy}
\author[b]{and A. Zichichi}

\affiliation[a]{CERN, Geneva, Switzerland}
\affiliation[b]{Dipartimento di Fisica e Astronomia dell'Universit\'a, Bologna, Italy}
\affiliation[c]{INFN, Bologna, Italy}
\affiliation[d]{Department of
Electronics and Communications, Politecnico di Torino and INFN Torino, Italy}
\affiliation[e]{Museo Storico della Fisica e Centro Studi e Ricercahe E. Fermi, Roma, Italy}
\affiliation[f]{Dipartimento di Fisica dell'Universit\'a, Salerno, Italy}


\emailAdd{francesca.carnesecchi@cern.ch, gianpiero.vignola@desy.de }

\abstract{The direct response of Silicon PhotoMultipliers being traversed by a MIP charged particle have been studied in a systematic way for the first time. Using beam test data, time resolution and the crosstalk probability have been measured. A characterization of the SiPM by means of a laser beam is also reported. The results obtained for different sensors indicate a measured time resolution  around 40-70 ps. Although particles are expected to traverse only one SPAD per event, crosstalk measurements on different sensors indicate  an unexpected higher value with respect to the one related to the sensor noise. }





\maketitle
\flushbottom

\section{Introduction}
\label{sec:intro}

Silicon PhotoMultipliers (SiPMs) are nowadays the photon detectors more widely used for  applications with scintillators (solid or liquid) or imaging, with important industrial impacts.  Their high detection efficiency (w.r.t. standard PhotoMultipliers), insensitivity to magnetic fields, easiness of handling and mechanical integration and low cost make their use particularly attractive. Many tests demonstrated also the excellent timing capabilities of such devices, reaching a time resolution of a few tens of ps for particle detection with scintillators \cite{cms,2020carnesecchi} and the possibility to go even lower as  demonstrated for single photons \cite{2020Gundacker, 2016Nemallapudi}.

Usually, the SiPMs are classified as detectors insensitive to charged particles, in the sense that such a signal would not affect the one coming from a certain number of photons. However, it has been pointed out that SiPMs can directly detect ionizing particles, without the need for a photon source like a scintillator: this has been observed with a relativistic ion beam \cite{2014Marrocchesi} or protons \cite{2014dascenzo} and studied using a single SPAD (Single Photon Avalanche Diode, i.e. a single pixel of a SiPM), in the last case with outstanding time resolution \cite{2021Gramuglia}. Recently the charged particle detection with SPADs was also studied with a simulation  \cite{2021Riegler}.  This possibility would allow their use as important elements for Particle IDentification (PID) in future experiments like time of flight applications (requiring a good tracking capability associated with good timing)  or as sensors for RICH applications combining its capabilities to detect photons and determine the particle impact point.  However, such possibility needs a more detailed understanding of the contribution of SPADs to the SiPM total output signal, whose amplitude depends on the number of fired SPADs.

A preliminary study of the response of SiPMs to the passage of charged particles was performed with Cosmic Rays in the INFN Bologna laboratories and demonstrated the possibility to have good signals \cite{Vignola}. Using a telescope with three equal SiPMs,  time resolutions around 140 ps were obtained; moreover, a crosstalk probability higher than that from the noise counts was measured. However, the low statistics available, spanning over several months, left several uncertainties in the results.\\
In the present work, the study was extended using  beam test data collected at the T10 facility of CERN PS. Several sensors were investigated  using a telescope with LGADs detectors \cite{2014Lgad} to define the track. Bench studies performed with a laser beam directly shoot on the sensors allowed also a better characterization of SiPM properties in terms of single SPAD response. 
The present study is focused on  time resolution and crosstalk measurements, reporting for the latter also a different and not explained higher value in the case of charged particles with respect  to the result obtained for the standard noise. 

The paper is organized as follow: after a description of the sensors and of the beam test and Laser setups in section~\ref{sec:setup}, section~\ref{sec:results} will describe the data analysis and the main results obtained.  In the conclusions, a few general remarks and prospects will be discussed.
\section{Experimental setups and preliminary measurements} 
\label{sec:setup}

\subsection{Detectors} 
\label{sec:detectors}
SiPMs with different characteristics and from different producers have been studied. 
The S13360-3050VE (henceforth called HPK3x3) produced by HPK has an active area of 3 $\times$ 3 mm$^2$ and a 50 $\mu$m square pixel pitch; for this sensor, a deeper study has been performed  with a laser beam to characterize the single SPAD response.
In addition, also NUV-HD-RH sensors prototypes produced by FBK have been studied; these detectors are based on the NUV-HD technology\cite{2019Gola} and have an active area of 1$\times$1 mm$^2$ and different hexagonal pixel pitches, henceforth called FBK1x1-15 and FBK1x1-20.

\begin{figure}[htbp]
\includegraphics [height=7.5 cm] {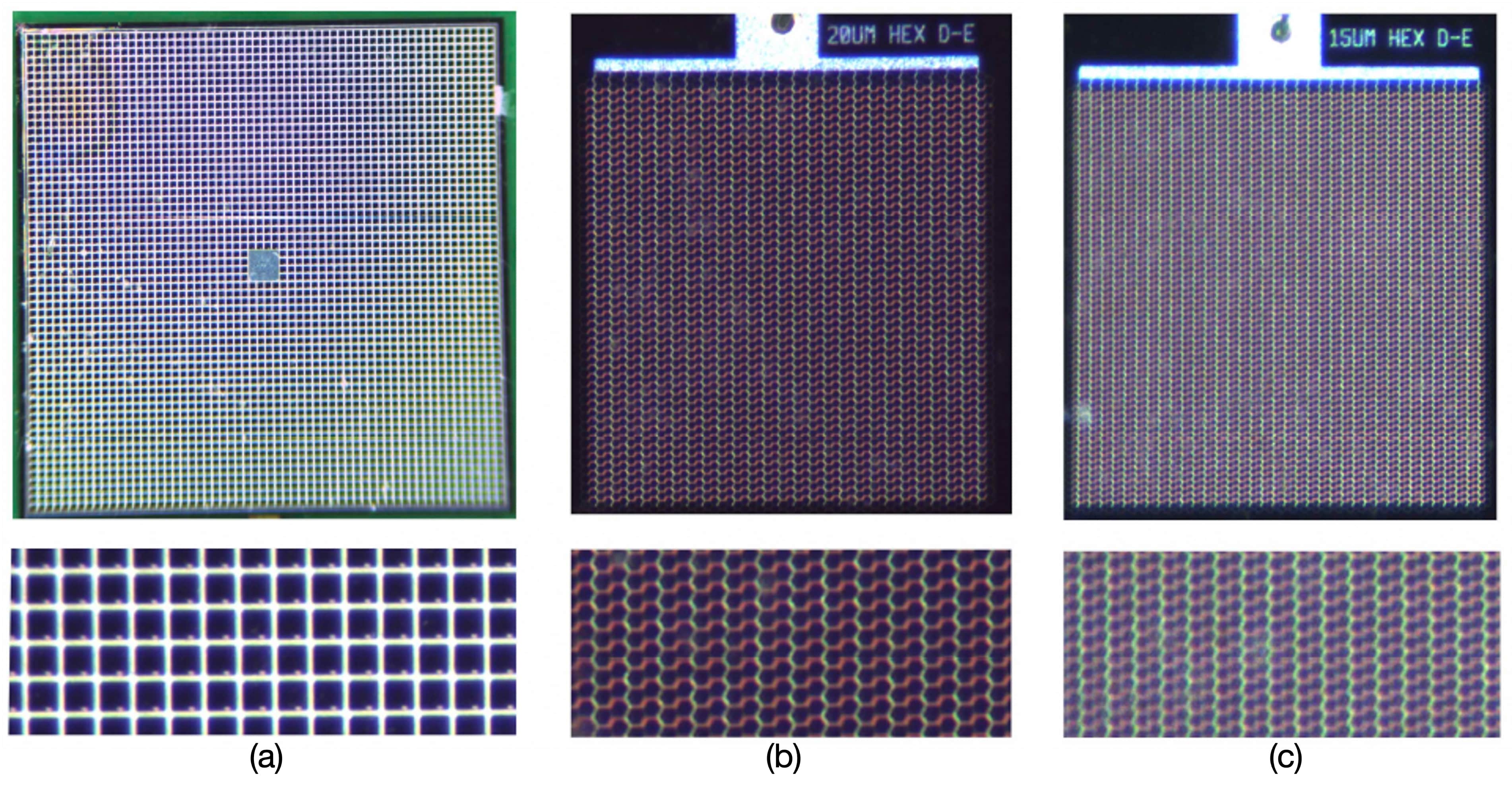}
\caption{Pictures of HPK3x3 (a), FBK1x1-20 (b) and FBK1x1-15 (c) SiPMs; in the bottom a magnified view of the SiPMs with the pixels subdivision and shapes visible.  }
\label{fig:photo_SiPM}
\end{figure}

In Table \ref{tab:sipm_char} the main characteristics of the photo-detectors are reported\footnote{\hyperlink{https://www.hamamatsu.com/resources/pdf/ssd/s13360-2050ve_etc_kapd1053e.pdf}{HPK S13360-3050VE datasheet}}, while Figure \ref{fig:photo_SiPM} shows the pictures of the three studied SiPMs.

\begin{table}[ht]
    \centering
\begin{tabular}{lccc}
\hline
Parameter & HPK3x3 & FBK1x1-20 & FBK1x1-15 \\
\hline
Effective area ($\text{mm}^2$) & 3 $\times$ 3  & 1$\times$1  & 1$\times$1 \\
Pixel pitch ($\mu \text{m}$) & 50  & 20 & 15\\
Number of pixels & 3584 &  2444 &   4464 \\
Fill factor ($\%$) & 74    & 72 & 63 \\
V$_{bd}$ (V) & 51.4 $\pm$ 0.1   & 32.9 $\pm$ 0.1 & 33.0 $\pm$0.1 \\
\hline
\end{tabular}
  \caption{Main characteristics of the SiPMs under test.} \label{tab:sipm_char}
\end{table}

The breakdown voltage V$_{bd}$ reported in Table \ref{tab:sipm_char} has been extracted from the measurements reported in \cite{2020carnesecchi} for the HPK3x3 and from I-V measurements for the FBK SiPMs using the method of ILD and LD \cite{2017Klanner}.


\subsection{Beam test setup} 
\label{subsec:tb}
The SiPMs response has been studied with MIPs at the T10 beamline of CERN-PS in November 2021. The beam was mainly composed of protons and pions with a momentum of 12 GeV/c. A small fraction of the data have been taken with a momentum of 2 GeV/c and a beam with increased positron content ($\sim20\%$).

\begin{figure}[htbp]
        \centering%
        \subfigure[\label{fig:SiPM-setup}]%
          {\raisebox{0.9cm}{\includegraphics [height=7cm] {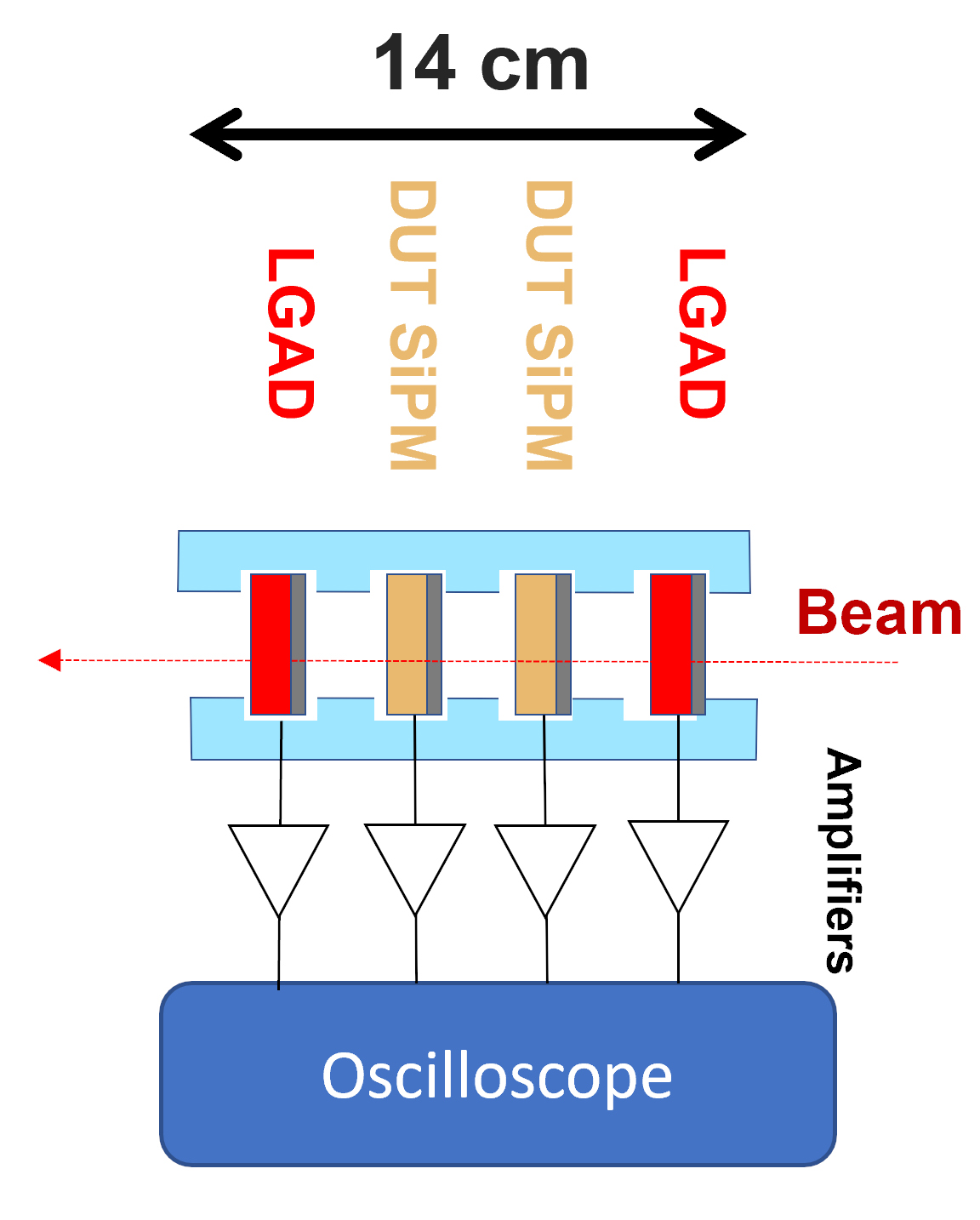}}}\quad
        \centering%
        \subfigure[\label{fig:SiPM-Signal}]%
          {	\includegraphics[height=8.5cm]{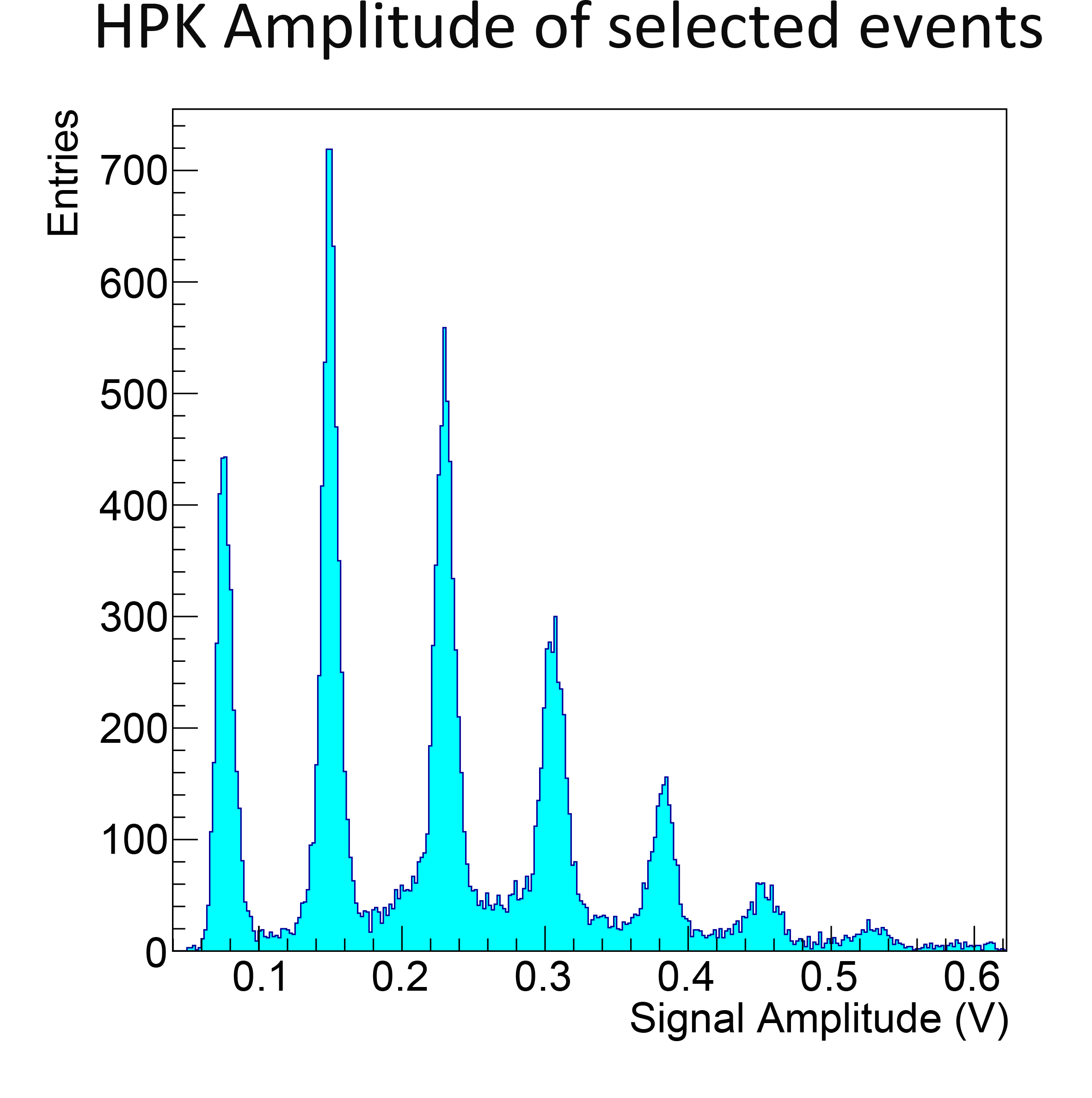}
	  }
        \caption{(a) Schematic representation of one configuration for the beam test setup with the LGAD sensors defining the track direction and the time zero t$_0$ for the event and the SiPMs Detectors Under Test (DUT). (b) Signals amplitude of HPK3x3 at OV 2.6 V after the cuts described in the text, the first peak corresponding to one SPAD.}
\end{figure}
The telescope was composed of the SiPMs under test plus  LGAD detectors (1x1 mm$^2$ area and 35 $\mu m$ or 25 $\mu m$ thick prototypes) \cite{LGAD} used as reference to track the beam particles and define the t$_0$ for the time resolution. An example of one setup is shown in Figure \ref{fig:SiPM-setup}, but the exact trigger and configuration changed from run to run due to different tested sensors. The whole setup was enclosed in a dark environment box at room temperature.

The SiPMs signals were independently amplified by 40 dB using  Cividec amplifiers\footnote{Cividec datasheet: \href{https://cividec.at/electronics-C2.html}{https://cividec.at/electronics-C2.html}}. The trigger was defined as the coincidence of two or three detectors in the telescope. At each trigger, all waveforms were stored using a Lecroy Wave-Runner 9404M-MS digital oscilloscope\footnote{Lecroy WaveRunner datasheet:

\href{https://teledynelecroy.com/oscilloscope/waverunner-9000-oscilloscopes/waverunner-9404m-ms}{https://teledynelecroy.com/oscilloscope/waverunner-9000-oscilloscopes/waverunner-9404m-ms}}. For the final offline analysis, the oscilloscope bandwidth was set to 1 GHz. 

\subsection{Signal selection}
\label{subsec:signal}
To evaluate the time resolution of the studied SiPMs, a set of cuts was applied in order to select only signal events and distinguish them from noise.
First of all the DCR (Dark Count Rate)  has been evaluated in the region before the trigger accepted time interval in order to determine the possible contamination of noise events in the signal region. 
Then given the LGADs trigger condition (t$_0$) only events with a DUT signal in a window of $\pm$ 2 ns from the trigger have been selected.
A further cut was applied on the DUT signals by removing the few events with important residuals of previous signals (DCR or MIP ) in a time window of 8 ns before the signal zone (-10 to -2 ns from the trigger).

An example of DUT signal amplitude distribution after amplification and selection is shown in Figure \ref{fig:SiPM-Signal} for the HPK3x3 sensor at an OverVoltage (OV, defined as the difference between the applied and the breakdown voltage) of 2.6V: it is possible to clearly distinguish signals due to single SPAD events (first peak of about 80 mV) and signals due to multiple cells events.

As a conclusion, in the worst case (high overvoltage) the contamination due to DCR was estimated to be less than 6$\%$ (less than 3$\%$ for HPK3x3) of the selected events, uniformly distributed along the selected window.

After the data selection, two options were pursued for the evaluation of the time of arrival of the signals: Constant Fraction Discrimination (CFD) and FIX Threshold analysis (FT). The latter measurements were corrected for time slewing using the signal amplitude. Notice that this correction was not possible for the SiPM signals that saturated the amplitude of the oscilloscope, nevertheless for such high signals the correction is negligible. 
The CFD and FT methods have been initially compared and resulted in a totally compatible time resolution.
The 40$\%$ CFD threshold was used for laser studies where we deal with single SPAD signals,  while an FT of approximately 30$\%$ of the single SPAD signal amplitude was used on the beam test data to allow the inclusion in the analysis of the saturated SiPM signals.

\subsection{Preliminary measurements with laser}
\label{subsec:preliminary}
A comprehensive characterization of the HPK3x3 has been performed using a laser beam impinging perpendicularly on the sensor, to measure the possible single SPAD contributions to the time resolution measurements. The Laser setup was developed in the laboratory of the INFN unit in Bologna during the preparation phase of the beam measurements \cite{Vignola}.

A picosecond pulsed 1054 nm laser (PiLas PiL036XSM) has been used; the light was then transmitted via a single-mode optical fiber and collimated and focused in a micrometric spot using Schäfter+Kirchhoff lens. 
In order to allow the correct focusing of the laser spot on the DUT a micro-moving device (manually controlled) has been used for the laser apparatus.
The DUT has in turn been moved with respect to the laser beam using a two axes Micrometer Positioning Stages (MPS) controlled by a LABVIEW program and capable of 1 micron precision steps. The preamplified signals were sent to a digital oscilloscope and  recorded. MPS and sensor (with amplifier) were placed inside a dark box.

\begin{figure}[htbp]
        \centering%
        \subfigure[\label{fig:SCAN_a}]%
          {\includegraphics [height=4.35cm] {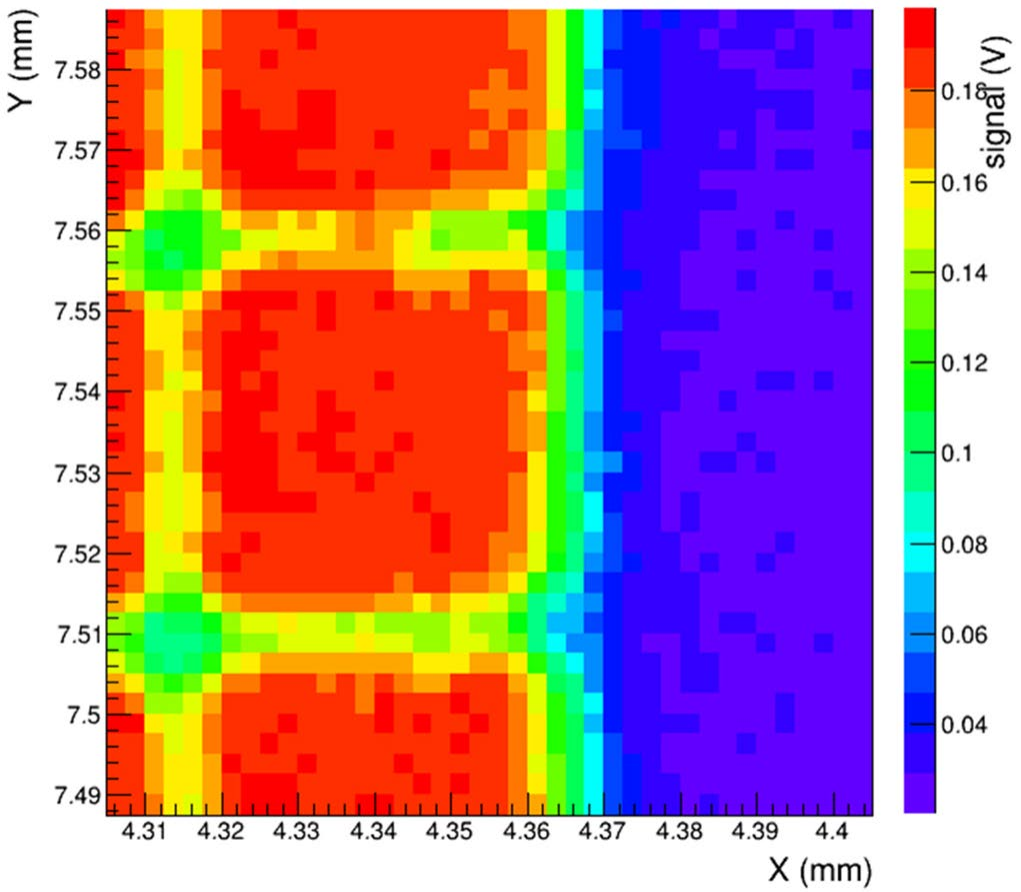}}\quad
        \centering%
        \subfigure[\label{fig:SCAN_b}]%
          {\raisebox{-0.0cm}{
          	\includegraphics[height=4.35cm]{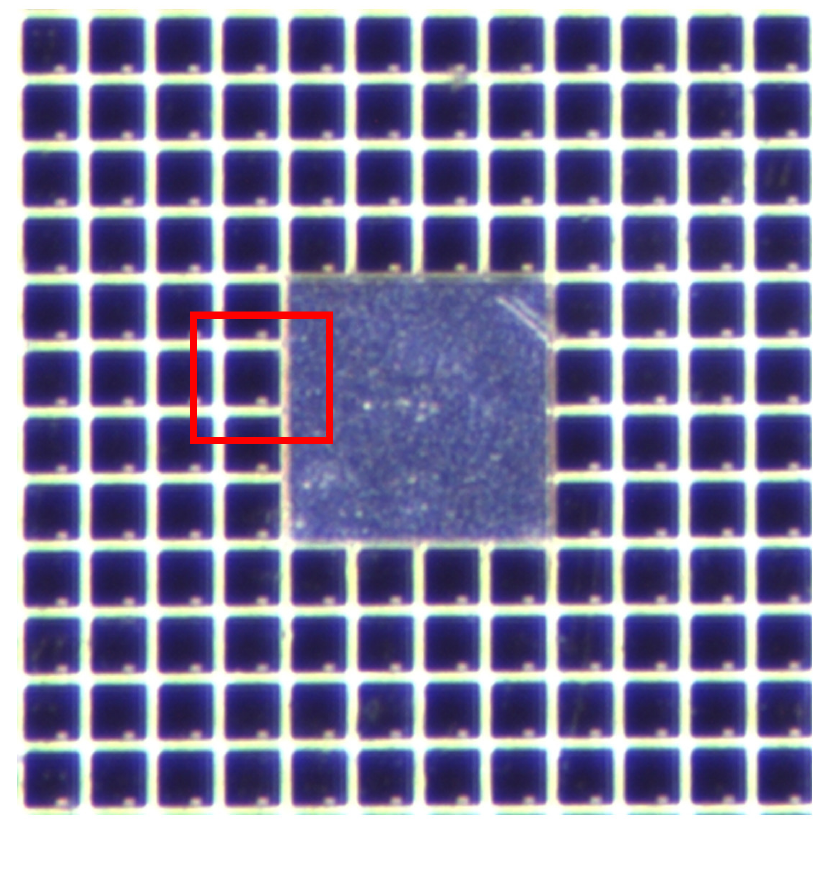}}}
          	\centering%
        \subfigure[\label{fig:SCAN_c}]%
          {\raisebox{-0.0cm}{
          	\includegraphics[height=4.35cm]{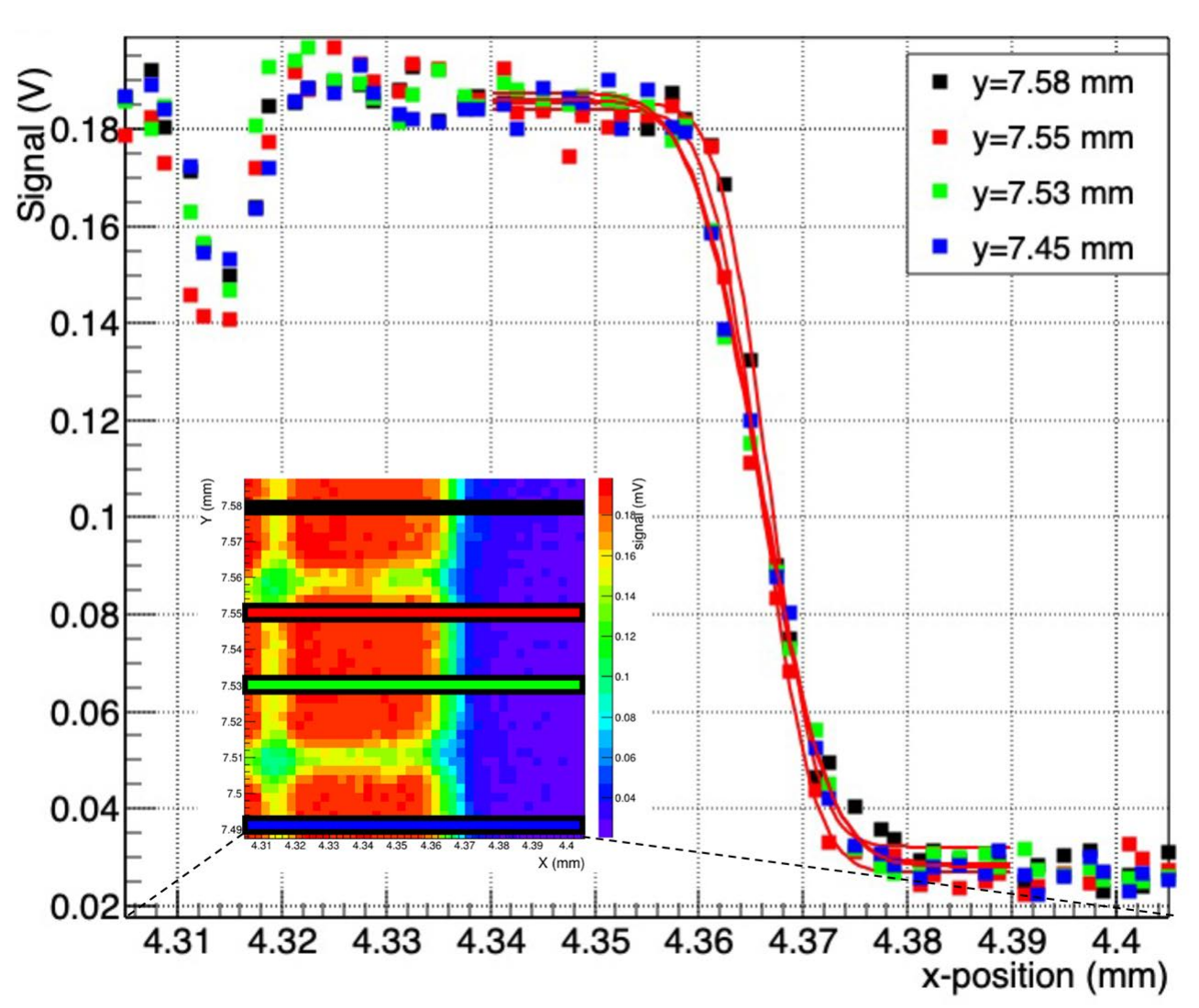}}
	  }
        \caption{Study of the HPK3x3 SiPM at 5.3 V of overvoltage using the laser setup. Detector response as a function of laser position (a). Photograph of the studied area (b). Selection of horizontal scans used to determine the laser spot size (c).}
        \label{fig:scan}
\end{figure}

It was then possible to study the HPK3x3 SiPM at the level of the single SPAD, measuring the laser spot size from the blind region at the center of the sensor (due to Through Silicon Via connections). 
Figure \ref{fig:SCAN_a} shows a SiPM sensitivity map in a particular transition region between the active area and the blind area (see Figure \ref{fig:SCAN_b}): the color panel in Figure \ref{fig:SCAN_a} indicates the average, over about 1000 laser pulses, signal amplitude in V. As shown in Figure \ref{fig:SCAN_c}, by performing horizontal scans in this region, it was possible to determine the size of the Gaussian shaped laser spot, fitting with a convolution of a step function and a Gaussian function (which respectively represent the transition between the active and dead region of the sensor and the shape of the laser spot). The focused laser spot was estimated to be about 10 µm diameter.
This value is below the SPAD pitch (50 µm) but not small enough to appreciate the dead area between adjacent cells (that, accordingly to  Table \ref{tab:sipm_char}, should be 13 $\mu$m); in fact, between SPADs, the sensor response  is only slightly lower than in the central areas of the cells. Note that with this laser spot diameter it was not possible to perform the same studies on the FBK sensors due to the smaller SPAD sizes. 
\\

\subsubsection{Time resolution of SiPM}
\label{subsubsec:timeResLaser}
The measured time resolution of a SiPM, $\sigma_\text{SiPM}$, is the result of several contributions and can be expressed as:
{
\begin{equation}
\sigma_\text{SiPM}^2=\sigma_\text{Intrinsic}^2+\sigma_\text{Jitter}^2=\sigma_\text{SPAD}^2+\sigma_\text{Delay}^2+\sigma_\text{Uniformity}^2+\sigma_\text{Jitter}^2
\label{eq:TR_sipm}\end{equation}
}
, the contribution from the readout electronics being totally negligible since it amounts to $\sim$ 7 ps using the WaveMaster SDA 816Zi-A oscilloscope\footnote{LeCroy WaveMaster datasheet: \href{https://docs.rs-online.com/035e/0900766b8127e31c.pdf}{https://docs.rs-online.com/035e/0900766b8127e31c.pdf}}.  $\sigma_\text{Jitter}$ is the jitter term due to interconnection between the detector and the amplifier and $\sigma_\text{Intrinsic}$ the intrinsic resolution of the detector. 
The latter includes $\sigma_\text{SPAD}$, the time resolution of a single SPAD, $\sigma_\text{Delay}$, the contribution due to the different signal collection times, depending on the position of the SPAD inside the sensor, $\sigma_\text{Uniformity}$, the spread due to SPAD-to-SPAD performances variation. 

All the listed contributions that lead to a worsening of the time resolution have been evaluated and reported in the following.

\subsubsection{SPAD time resolution and jitter}
\label{subsubsec:timeResLaser2}
The precision of the laser spot allowed to perform some preliminary measurements of the time resolution of single SPADs.

In order to select the data and distinguish them from noise, the same cuts were applied on signal amplitude, shape \cite{2016deriv} and signal position as reported in Sec.\ref{subsec:signal}. For this analysis a further cut was introduced: in order to evaluate just the response of the single SPAD reducing the possibility of the laser halo impinging on a nearby SPAD, events with signal amplitude corresponding to 2 or more SPADs have been excluded. 
The events passing the cuts have been analysed considering the time difference with respect to the t$_0$ given from the laser itself (which had a negligible jitter).

The time distribution obtained has then been fitted using a q-Gaussian  function with a measured sigma ($\sigma_\text{SiPM}$). 
The analysis has been repeated for several values of the CFD. 
In addition, the  electronic noise jitter contribution has been evaluated \cite{2014Acerbi}; a time window ahead the trigger region has been chosen, in order to select only the part of the waveforms which does not contain the signal event itself. In this way it was possible to evaluate the RMS noise. The jitter has then been subtracted from $\sigma_\text{SiPM}$ to obtain the final intrinsic time resolution $\sigma_\text{Intrinsic}$ (see Eq. \ref{eq:TR_sipm}). Since the laser was focused on a single SPAD, the intrinsic time resolution here measured corresponds to the single SPAD time resolution $\sigma_\text{SPAD}$.

\begin{figure}[h]
        \centering%
          \includegraphics [width=9cm]{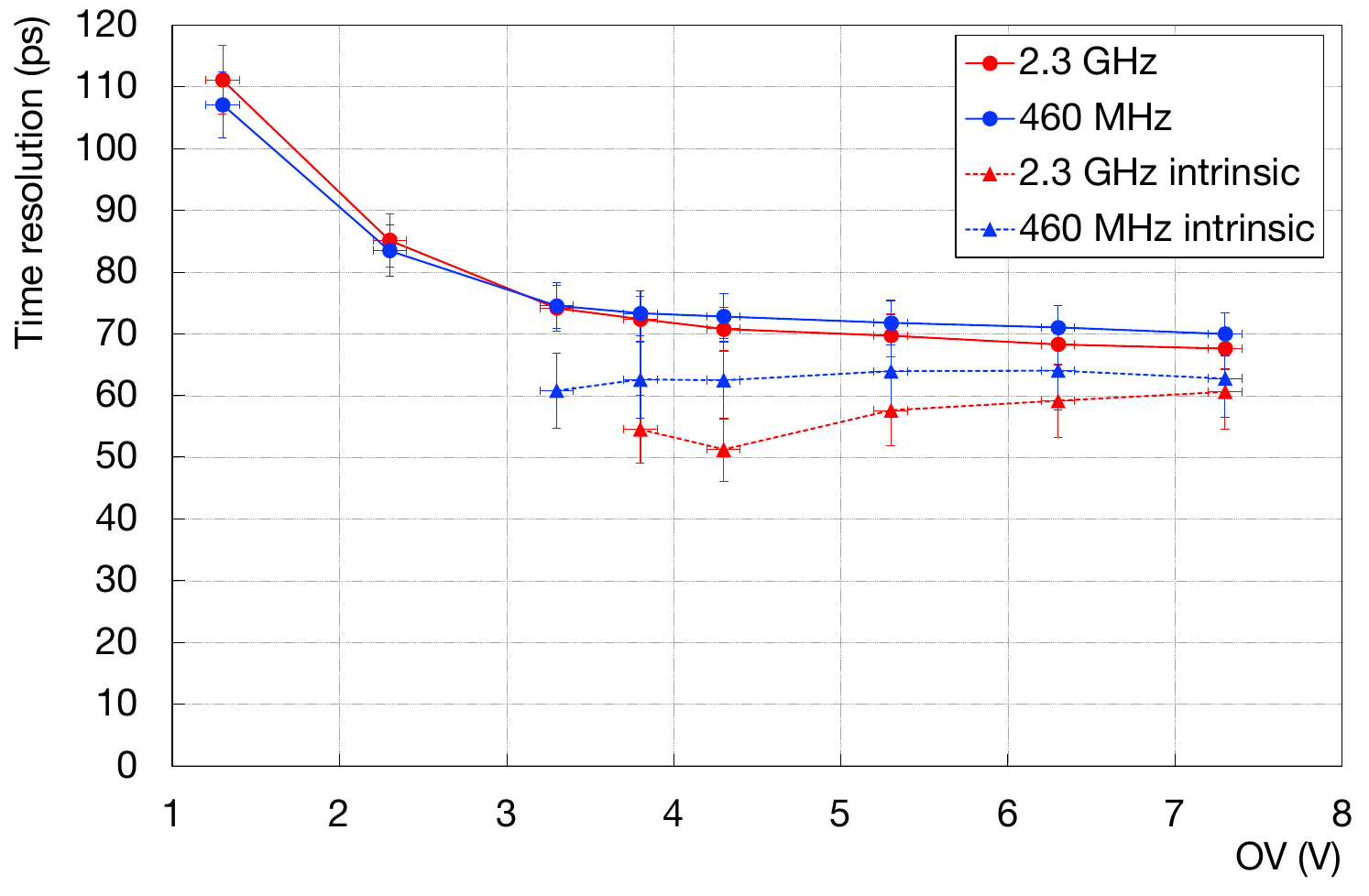}
        \caption{Laser test: measured and intrinsic time resolution of HPK3x3 versus the overvoltage for two bandwidths. A CFD=40$\%$ was used.}
          \label{fig:Lasres}
\end{figure}

In Figure \ref{fig:Lasres} the measured and the SPAD intrinsic time resolution versus the OV is reported for a CFD=40\% and for a bandwidth of the oscilloscope equal to 460 MHz and 2.3 GHz. The measured values for the two bandwidths are compatible within errors and demonstrate that the 1 GHz value used for the final beam test analysis gives stable results. 
As expected, for higher voltage the time resolution improves and reaches a plateau around 4.5 V OV: notice that a higher value of voltage corresponds to a higher detector gain. 

Concerning the intrinsic resolutions there is a tendency of smaller values for the 2.3 GHz, but still compatible within errors. In principle the two bandwidths should give similar results, but the subtraction of $\sigma_\text{Jitter}$ has been particularly challenging because of the important electronic noise that contributes in a slightly different way depending on the bandwidths. Considering that only events with one fired SPAD have been considered, the same problem affects in a stronger way the lower OV region, where the S/N ratio is smaller. For this reason, in the first two points it was not possible to subtract it reliably.
As a consequence, for the results in Section~\ref{sec:results}, only the measured values and not the intrinsic ones will be reported. 

As a final remark, for the HPK3x3 SiPM and at OV larger than 4.5 V, the measured intrinsic time resolution of the single SPAD is around 60 ps (corresponding to a measured time resolution of $\sim$ 69 ps), compatible with the one obtained in \cite{2020Gundacker} for a single photon time resolution study on the same kind of SiPM.

\subsubsection{SPAD uniformity and Delay}\label{par:DiffSPADs}
Using 6.3 V overvoltage value, the evaluation of the time resolution in various SPADs of the sensor has been carried out. As detailed in \cite{Vignola}, nine SPADs distributed uniformly on the sensor were chosen, intentionally involving potentially critical areas such as edges or center of the sensor, in order to maximize any discrepancy between the values. It was found that the measured time resolution of the single SPAD is almost uniform within the sensor, resulting in a negligible contribution due to this factor,  $\sigma_\text{Uniformity}\sim 
5ps$.

Focusing instead on the delay in the response of the SPADs with respect to the trigger given by the laser, some differences were observed in the different areas of the sensor. A maximum difference of approximately 80 ps was observed between the selected SPADs. This corresponds to a $\sigma_{Delay}$ of $\sim$ 25 ps; this result is consistent with those of other studies performed on similar SiPMs \cite{2016Nemallapudi}.

As a conclusion, in a first approximation, using Eq.\ref{eq:TR_sipm} and adding all contributions, a time resolution $\sim$ 74 ps  for the $\sigma_\text{SiPM}$ of HPK3x3 is expected for an OV of 6.3 V. 
\section{Beam test results} 
\label{sec:results}
In this section, details of the analysis of beam test data are reported.
The results on time resolution and crosstalk are discussed. The signal selection has been already described in Section \ref{subsec:signal}.

The data analysis for timing measurements is similar to the one used for the laser setup (see Section \ref{subsubsec:timeResLaser} and \ref{subsubsec:timeResLaser2}) with some differences.
In this case, the time difference between the SiPM under study and the LGADs has always been considered.
 \begin{figure}[h!]
        \centering%
          {\includegraphics [width=12cm]{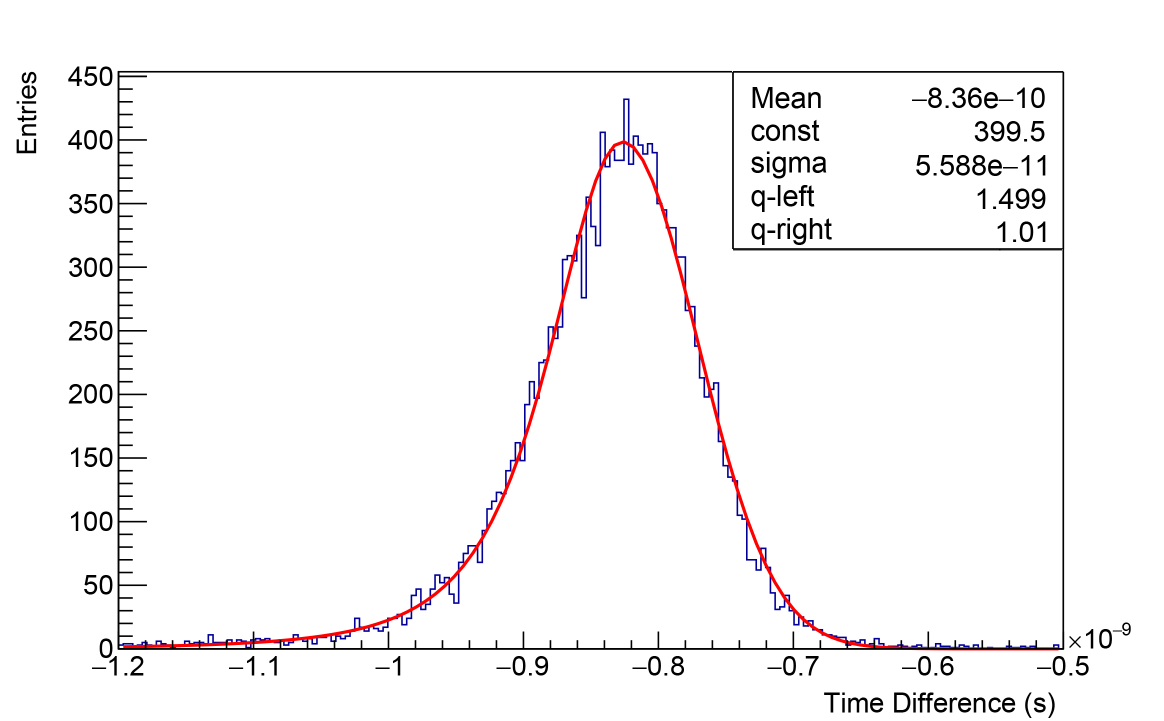}}
        \caption{Time difference between LGAD and FBK1x1-20 at 6.1 OV after the time slewing correction. The distribution has been fitted with a q-Gaussian.}
          \label{fig:timediff}
\end{figure}
The distribution has been fitted with a q-Gaussian to take
into account also the tails; the measured parameter ($\sigma$) has then been used to extract the final measured time resolution: $\sigma_\text{measured}=\sqrt{\sigma^2-\sigma_\text{LGAD}^2}$, where $\sigma_\text{LGAD}$=26-36 ps depending on the voltage applied and on LGAD used as reference \cite{LGAD}. An example of the measured time distribution is reported in Figure \ref{fig:timediff} for the FBK1x1-20 sensor.

\subsection{Results on efficiency and crosstalk}
\label{subsec:result-crosstalk}
A measurement of the efficiency of HPK3x3 has been done with beam test data selecting events with a coincidence between the upstream LGAD and the downstream FBK1x1-20: the trigger region was much smaller than the HPK3x3 area. The measured efficiency was found to be $\sim$95 $\%$ for all OV tested.
The value is higher than what expected from the fill factor (see Table\ref{tab:sipm_char}). This would make a SiPM a much more attractive technology as a tracking detector. 
 \begin{figure}[h!]
        \centering%
          {\includegraphics [width=12cm]{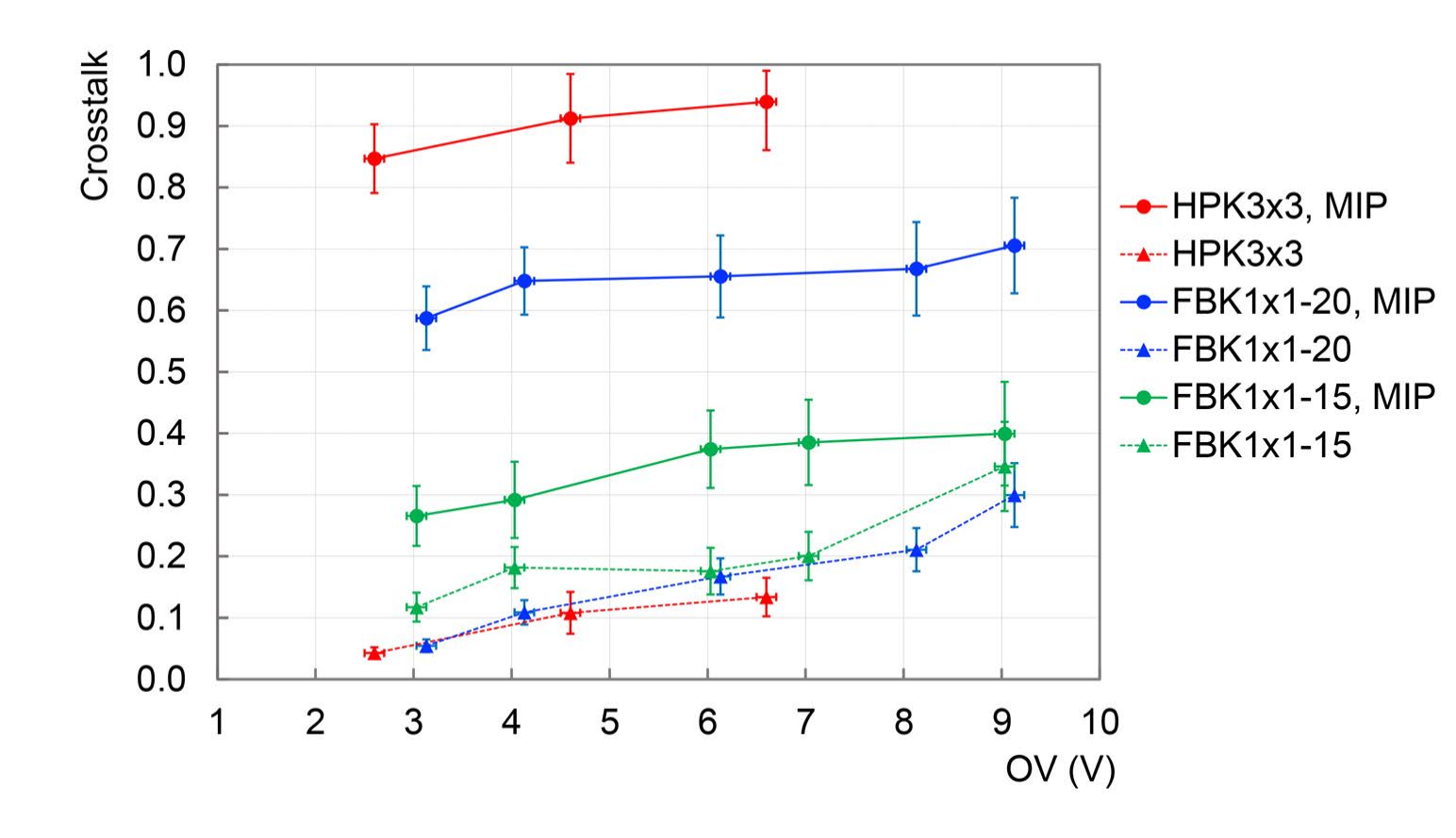}}
        \caption{Measured crosstalk fraction versus overvoltage for beam test data (MIP, full lines) and standard evaluation via DCR (dotted lines).}
          \label{fig:crosstalk}
\end{figure}

The above result suggested to investigate also the  crosstalk probability between SPADs and a further analysis was performed  with beam test data  measuring the number of MIP events with 1, 2, ... n fired SPADs. 
Usually, such distribution is used to study the intrinsic  crosstalk of the SiPM, defined starting from the DCR signals as:
\begin{equation}
\text{Crosstalk} = \frac{\text{DCR with}  \geq 2 \text{ pixels firing}}{\text{DCR with}  \geq 1 \text{ pixels firing}}
\end{equation}

Notice that usually such crosstalk is related to nearby fired SPADs with almost super-imposed signals, while the probability of two random SPADs firing within the same time interval is negligible. In Figure \ref{fig:crosstalk} the standard crosstalk measured from  DCR (evaluated in a region before the trigger to be sure that possible effect related to the beam spill were included) for the detectors under test is reported (dotted lines). It increases with the overvoltage, reaching a maximum of $\sim$ 20-30 $\%$ at high OV. 

The same analysis has been applied in the trigger region. Indeed, a binary response is expected from the single SPAD, independently from being hit by one or more photons or by a MIP.
Moreover, the geometry of the telescopes and the coincidence logic makes very improbable to have inclined tracks crossing more than one SPAD. 
So for a MIP traversing a SiPM it could be expected a behaviour similar to standard crosstalk.  

In Figure \ref{fig:crosstalk} the crosstalk measured with MIP particles (continuous lines) is compared with the standard DCR one (dotted lines). As can be noticed, a much higher value is observed for particles traversing the SiPM for all sensor types and for all the over-voltages.
Notice that the HPK3x3 result was obtained with a beam  with a higher contribution from positrons.  A similar effect was observed in \cite{2014Marrocchesi} with ion beams. 

These results would require the presence of a mechanism that induces a signal on nearby SPADs either in the detector package or in the silicon sensor itself and seem to depend on the details of the manufacturing process.
Further investigations on the possible underlying mechanisms are needed. Indeed a single MIP higher induced crosstalk would be beneficial in terms of time resolution, efficiency and event discrimination.

\subsection{Results on time resolution}
\label{subsec:result-time}
\begin{figure}[h!!]
        \centering%
        \subfigure[\label{fig:CR_timeRes}]%
          {\includegraphics [height=5.6cm] {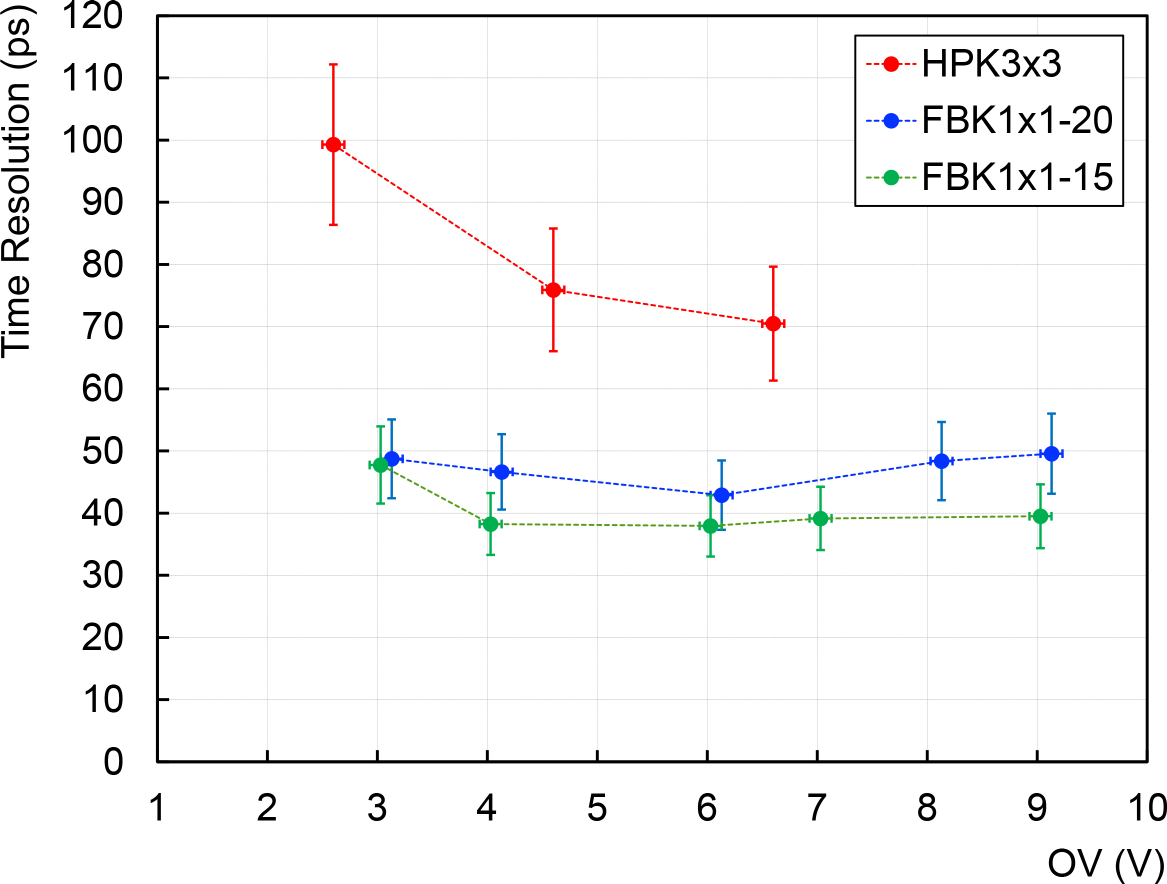}}
        \centering%
        \subfigure[\label{fig:timRes_SPAD}]%
          {	\includegraphics[height=5.6cm]{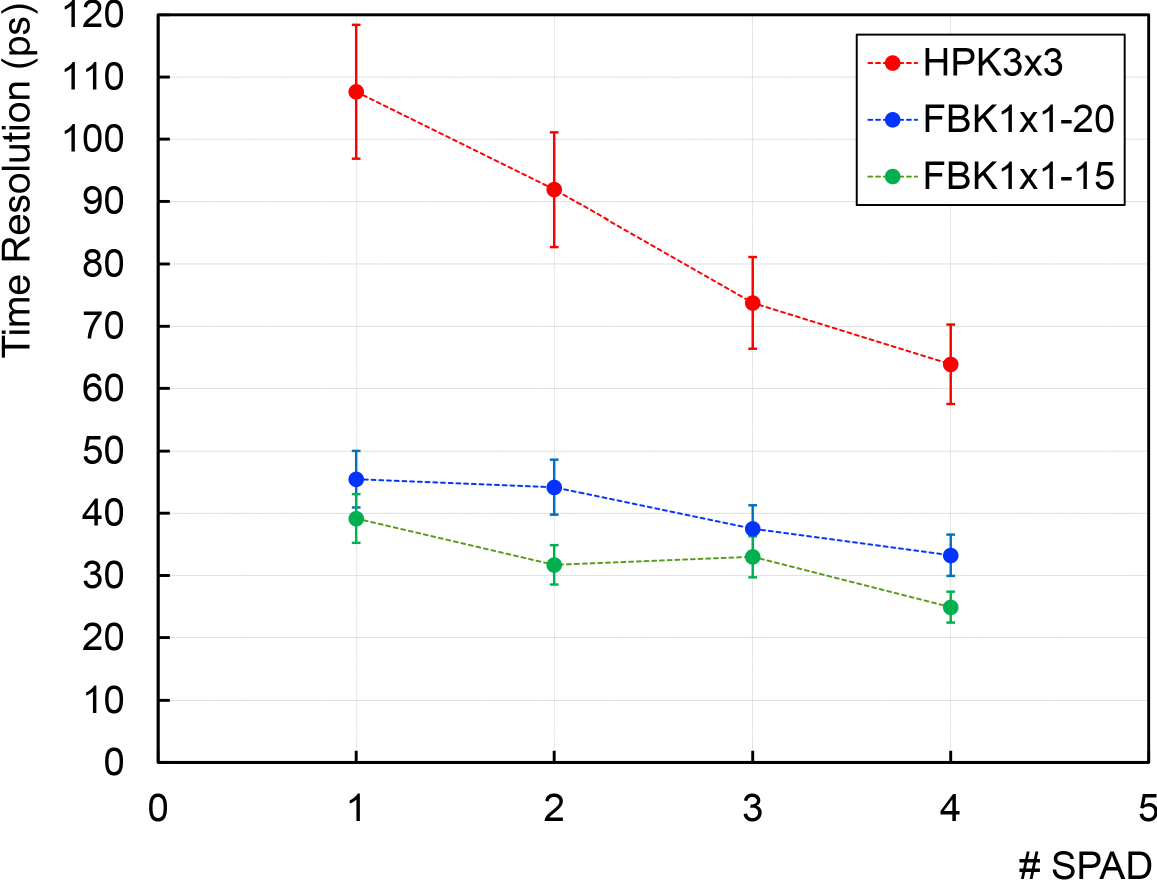}
	  }
        \caption{Time resolution obtained in the beam test setups: (a) versus overvoltage and (b)  at an OV of around 6 V as a function of the number of SPADs fired, with 4 indicating $\geq$ 4.}
\end{figure}

In Figure \ref{fig:CR_timeRes} the measured time resolution $\sigma_\text{SiPM}$ is reported as a function of the OV for the three sensors.
The measurement is done including all signals, independently of the number of fired SPADs. A better time resolution has been obtained for the FBK SiPMs, partially explained from the smaller area (lower capacitance).
For the HPK3x3 the time resolution improves with higher voltage;
for the FBK SiPMs the voltage dependence is less marked and the time resolution is almost constant within the errors.

In Table \ref{tab:res} the time resolution obtained at the highest OV for all the SiPMs are summarized.

\begin{table}[ht]
    \centering
\begin{tabular}{lccc}
\hline
 & OverVoltage  & Time resolution\\
\hline
HPK3x3  & (6.6 $\pm$ 0.1) V & (70 $\pm$ 9) ps\\
FBK1x1-20  & (9.1 $\pm$ 0.1) V & (50 $\pm$ 6) ps\\
FBK1x1-15  & (9.0 $\pm$ 0.1)V & (39 $\pm$ 5) ps\\
\hline
\end{tabular}
  \caption{Time resolution for HPK3x3, FBK1x1-15 and FBK1x1-20 for a given OV obtained in a beam test setup at room temperature.}
\label{tab:res}
\end{table}

In Figure \ref{fig:timRes_SPAD} the time resolutions as a function of the number of SPADs that fired for a single MIP event have been reported. Note that it is possible to measure up to 4 or more SPADs ( see also figure \ref{fig:SiPM-Signal}).
For a given number of SPAD the FBK sensors time resolution is always better than HPK3x3.
The improvement with increasing number of SPADs is evident. The trend is more marked for the HPK3x3, while it is flatter for the FBK sensors, indicating a more stable operation.
\section{Conclusions}
\label{sec:conclusions}
In this paper, systematic measurements on the response of SiPMs to the passage of MIP charged particles are reported for the first time. A bench test with a laser beam allowed a better characterization of the SiPM in terms of single SPAD contribution to the final results. Tests have been performed at the T10 CERN PS beamline. Using different sensors from different foundries, area and SPAD geometries, time resolutions in the range 40-70 ps have been achieved, including the electronic jitter contribution. A  measurement of the crosstalk fraction is also reported, resulting in a higher value for charged particles traversing the SiPM with respect to the standard one related to dark count rate. The results point to an efficiency higher than the simple fill factor of the devices. Both the higher crosstalk and efficiency would require the presence of a mechanism that induces a signal on nearby SPADs either in the detector package or in the silicon sensor itself. Future tests are needed to understand the origin of such effect and would require a strong interaction with the SiPM producers and a precise study of the particle energy and type dependence.

\acknowledgments

The authors wish to thank M. Basile for a careful reading of the manuscript and  A.Gola and A.Mazzi of FBK for useful discussions. The support of the mechanical and electronic workshops of the INFN Unit of Bologna has been highly appreciated. We thanks also the CERN-PS operator team for the support.


\end{document}